\documentclass[english,twoside,a4paper,10pt]{article}

\usepackage[latin1]{inputenc}
\usepackage[T1]{fontenc}
\usepackage{amsmath}
\usepackage{amsfonts}
\usepackage{graphicx}
\usepackage{subfigure}
\usepackage{a4wide}
\usepackage{amssymb}
\usepackage{fancyhdr}
\usepackage{mathrsfs}

\begin{document}

\title{\bf Non-singular exponential gravity: a simple theory for
early- and late-time accelerated expansion}
\author{
E.~Elizalde$^1$\footnote{E-mail address: elizalde@ieec.uab.es and elizalde@math.mit.edu},
S.~Nojiri$^{2,3}$\footnote{E-mail
address: snojiri@yukawa.kyoto-u.ac.jp and nojiri@cc.nda.ac.jp},
S.~D. Odintsov$^{1,4}$\footnote{E-mail address: odintsov@ieec.uab.es; also at Tomsk 
State Pedagogical University},
L.~Sebastiani$^5$\footnote{E-mail
address: l.sebastiani@science.unitn.it}
\ and S. Zerbini$^5$\footnote{E-mail
address: zerbini@science.unitn.it}\\
\\
\begin{small}
$^1$Consejo Superior de Investigaciones Cient\'{\i}ficas, ICE/CSIC-IEEC,
\end{small}\\
\begin{small}
Campus UAB, Facultat de Ci\`{e}ncies, Torre C5-Parell-2a pl, E-08193
Bellaterra (Barcelona) Spain
\end{small}\\
\begin{small}
$^2$Department of Physics, Nagoya University, Nagoya 464-8602,
Japan
\end{small}\\
\begin{small}
$^3$Kobayashi-Maskawa Institute for the Origin of Particles and the
Universe,
\end{small}\\
\begin{small}
Nagoya University, Nagoya 464-8602, Japan
\end{small}\\
\begin{small}
$^4$Instituci\'{o} Catalana de Recerca i Estudis Avan\c{c}ats
(ICREA)
\end{small}\\
\begin{small}
and Institut de Ci\`{e}ncies de l'Espai (IEEC-CSIC), Campus UAB, Facultat
de Ci\`{e}ncies,
\end{small}\\
\begin{small}
Torre C5-Parell-2a pl, E-08193 Bellaterra
(Barcelona), Spain
\end{small}\\
\begin{small}
$^5$Dipartimento di Fisica, Universit\`a di Trento
and Istituto Nazionale di Fisica Nucleare
\end{small}\\
\begin{small}
Gruppo Collegato di Trento, Italia
\end{small}\\
}
\date{}

\maketitle


\begin{abstract}

A theory of exponential modified gravity which explains both
early-time inflation and late-time acceleration, in a unified way, is proposed.
The theory successfully passes the local tests and fulfills the cosmological bounds and,
remarkably, the corresponding inflationary era is proven to be unstable.
Numerical investigation of its late-time evolution leads to the conclusion that the
corresponding dark energy epoch is not distinguishable from the one for the $\Lambda$CDM model.
Several versions of this exponential gravity, sharing similar properties, are formulated.
It is also shown that this theory is non-singular, being
protected against the formation of finite-time future singularities.
As a result, the corresponding future universe evolution asymptotically tends,
in a smooth way, to de Sitter space, which turns out to be the final attractor of the system.

\end{abstract}


\def\thesection{\Roman{section}}
\def\theequation{\Roman{section}.\arabic{equation}}

\section{Introduction \label{SectI}}

Modified gravity is getting a lot of attention from the scientific community, owing in particular
to the remarkable fact that it is able to describe early-time inflation as well as the late-time
(dark energy) acceleration epoch in a unified way. This approach appears to be very economical,
as it avoids the introduction of any extra dark component (inflaton 
or dark energy of any kind)
for the explanation of both inflationary epochs. Moreover, it may be expected that, with some additional effort,
it will be able to provide a reasonable resolution of the dark matter problem as well as of reheating,
two other important issues in the description of the evolution of our universe. Furthermore, as a by-product,
modified gravity has the potential to lead to a number of interesting applications in high-energy physics.
In particular, $R^2$ gravity, which provides a simple example of modified gravity, may serve for the unification of all fundamental interactions, including quantum gravity, in an asymptotically-free theory \cite{bos92}.  Modified gravity may also be used to provide the scenario for the resolution of the hierarchy problem of high energy physics \cite{cenoz0601}. Finally, the corresponding string/M-theory approach modifies gravity already in the low-energy effective-action approximation, so that a theory of the kind considered appears to be quite natural from very fundamental considerations
(see, for instance, \cite{Nojiri:2003rz}).

Presently, a number of viable $F(R)$ gravities leading to a unified
description as explained have been identified (for a recent review, see \cite{F(R)}, and for a
description of the observable consequences of such models, see \cite{F(R)a}).
It goes without saying that all those models are constrained to obey the known local tests,
as well as cosmological bounds. However, this might not be such a severe problem, since already
the first model proposed \cite{Nojiri2003} which unified inflation with dark energy already satisfied many of these local tests. The real internal problem of $F(R)$ gravity
is related with its being a higher-derivative theory, which renders it highly
non-trivial. This means that it is hard, in fact, to explicitly work with
such theories and to get observable predictions from them.

The main aim of this paper is to propose a reasonably simple but indeed
viable version of $F(R)$ gravity which consistently describes the unification of the
inflationary epoch with the dark energy stage, while satisfying the known local tests
and cosmological bounds. Specifically, to address the issues above, we here propose
exponential gravity, which on top of being simple is moreover free from any kind of
finite-time future singularity and exhibits other very interesting properties, as we will see.

The paper is organized as follows. In the next section we briefly review
$F(R)$ gravity as well as the corresponding FRW cosmological equations. Special
attention is paid to de Sitter and spherically-symmetric solutions. Sect.~\ref{SectIII} is
devoted to the discussion of the viability conditions in $F(R)$ gravity.
These conditions are investigated for the simple and realistic theory of exponential gravity,
proposed as a dark energy model, in Sect.~\ref{SectIV}. In Sect.~\ref{SectV} we carry out
a detailed analysis of our explicit proposal: exponential gravity which describes
in a natural, unifying way both early-time inflation and late-time acceleration. It is
there demonstrated, too, that the model leads to a satisfactory, graceful exit from inflation
(the de Sitter inflationary solution being unstable). In Sect.~\ref{SectVI} we show that the
theory does not lead to any sort of finite-time future singularities. A careful numerical
investigation of late-time cosmological dynamics is carried out in Sect.~\ref{SectVII}.
It will be demonstrated there that exponential gravity makes specific predictions which are
not distinguishable from those of the $\Lambda$CDM model in the dark energy regime.
The asymptotic behavior of the theory at late times is investigated in Sect.~\ref{SectVIII}.
Section \ref{detailed} is devoted to a somehow different model, a variant of exponential gravity which unifies unstable inflation with the dark energy epoch and which is protected against future
singularities by construction. This opens the window to other variations of the basic model sharing
all its good properties. In the discussion section \ref{SectX}, a final summary and outlook are provided, and there is an Appendix on the Einstein frame.



\setcounter{equation}{0}

\section{$F(R)$-gravity: general overview and FRW cosmology
\label{SectII}}

\subsection{The classical action}

The action of modified $F(R)$ theories is \cite{F(R)}:
\begin{equation}
S=\int d^4x\sqrt{-g}\left[
\frac{F(R)}{2\kappa^2}+\mathcal{L}^{\mathrm{(matter)}}\right]\,,\label{action}
\end{equation}
where $g$ is the determinant of the metric tensor $g_{\mu\nu}$,
$\mathcal{L}^{\mathrm{(matter)}}$ is the matter Lagrangian and $F(R)$ a generic function of the Ricci scalar, $R$.
In this paper we will use units where $k_\mathrm{B}=c=\hbar=1$ and denote the gravitational
constant $\kappa^2=8\pi G_N\equiv8\pi/M_\mathrm{Pl}^2$, with the Planck mass being
$M_\mathrm{Pl}=G^{-1/2}_N=1.2\times 10^{19}\text{GeV}$.
We shall write
\begin{equation}
F(R)=R+f(R)\,.
\end{equation}
The modification is represented by the function $f(R)$ added to the classical term $R$ of
the Einstein-Hilbert action of General Relativity (GR). In what follows we will analyze
modified gravity in this form, explicitly separating the contribution of GR from its modification.

By variation of Eq.~(\ref{action}) with respect to $g_{\mu\nu}$, we
obtain
the field equations:
\begin{equation}
R_{\mu\nu}-\frac{1}{2}Rg_{\mu\nu}=\kappa^2 \left
(T^{{\mathrm{MG}}}_{\mu\nu}
+\tilde{T}^{\mathrm{(matter)}}_{\mu\nu}\right) \,.\label{EqEinstMod}
\end{equation}
Here, $R_{\mu\nu}$ is the Ricci tensor and the part of modified
gravity is
formally included into the `modified gravity' stress-energy tensor
$T^{{\mathrm{MG}}}_{\mu\nu}$, given by
\begin{equation}
T_{\mu\nu}^{{\mathrm{MG}}}=\frac{1}{\kappa^2
F'(R)}\left\{\frac{1}{2}g_{\mu\nu}[F(R)-RF'(R)]
+(\nabla_{\mu}\nabla_{\nu}-g_{\mu\nu}\Box)F'(R)\right\}\,.
\end{equation}
The prime denotes derivative with respect to the curvature $R$,
$\nabla_{\mu}$
is the covariant derivative operator associated with $g_{\mu\nu}$ and
$\Box\phi\equiv g^{\mu\nu}\nabla_{\mu}\nabla_{\nu}\phi$ is the
d'Alembertian for a scalar field $\phi$.
$\tilde{T}^{\mathrm{(matter)}}_{\mu\nu}$ is given by the non-minimal
coupling
of the ordinary matter stress-energy tensor
$T^{\mathrm{(matter)}}_{\mu\nu}$
with geometry, namely,
\begin{equation}
\tilde{T}^{\mathrm{(matter)}}_{\mu\nu}=\frac{1}{F'(R)}T^{\mathrm{(matter)}}_
{\mu\nu}\,.
\end{equation}
In general, $T^{\mathrm{(matter)}}_{\mu\nu}=\text{diag}(\rho,p,p,p)$,
where
$\rho$ and $p$ are, respectively, the matter energy-density and
pressure.
When $F(R)=R$, $T^{{\mathrm{MG}}}_{\mu\nu}=0$ and
$\tilde{T}^{\mathrm{(matter)}}_{\mu\nu}=T^{\mathrm{(matter)}}_{\mu\nu}$.

It should be noted that, due to the diffeomorphism invariance of the total action, only
$T^{\mathrm{(matter)}}_{\mu\nu}$ is covariantly conserved and, formally, $\frac{\kappa^2}{F'(R)}$
may be interpreted as an effective gravitational constant, assuming we are dealing with models such that $F'(R)>0$.

The trace of Eq.~(\ref{EqEinstMod}) reads
\begin{equation}
3\Box F'(R)+RF'(R)-2F(R)=\kappa^2 T^{\mathrm{(matter)}}\,, \label{scalaroneq}
\end{equation}
with $T^{\mathrm{(matter)}}$ the trace of the matter stress-energy tensor.
We can rewrite this equation as
\begin{equation}
\Box F'(R)=\frac{\partial V_{\mathrm{eff}}}{\partial F'(R)}\,,
\end{equation}
where
\begin{equation}
\frac{\partial V_{\mathrm{eff}}}{\partial
F'(R)}=\frac{1}{3}\left[2F(R)-RF'(R)+\kappa^2 T^{\mathrm{(matter)}}\right]\,,
\end{equation}
$F'(R)$ being the so-called `scalaron' or the effective scalar degree
of freedom.
On the critical points of the
theory, the effective potential $V_{\mathrm{eff}}$ has a maximum (or
minimum), so that
\begin{equation}
\Box F'(R_{\mathrm{CP}})=0\,,\label{criticalpoint1}
\end{equation}
and
\begin{equation}
2F(R_{\mathrm{CP}})-R_{\mathrm{CP}}F'(R_{\mathrm{CP}})=-\kappa^2
T^{\mathrm{(matter)}}
\,.\label{criticalpoint2}
\end{equation}
Here, $R_{\mathrm{CP}}$ is the curvature of the critical point.
For example, in absence of matter, i.e. $T^{\mathrm{(matter)}}=0$, one has the de Sitter critical point
associated with a constant scalar curvature $R_{\mathrm{dS}} $, such that
\begin{equation}
2F(R_{\mathrm{dS}})-R_{\mathrm{dS}}F'(R_{\mathrm{dS}})=0\,.\label{dScondition}
\end{equation}

Performing the variation of Eq.~(\ref{scalaroneq}) with respect to $R$,
by
evaluating $\Box F'(R)$ as
\begin{equation}
\Box F'(R)=F''(R)\Box R+F'''\nabla^{\mu} R\nabla_{\nu} R\,,
\end{equation}
we find, to first order in $\delta R$,
\begin{eqnarray}
&& \Box
R+\frac{F'''(R)}{F''(R)}g^{\mu\nu}\nabla_{\mu}R\nabla_{\nu}R
 -\frac{1}{3F''(R)}
\left[2F(R)-RF'(R)+\kappa^2 T^{\mathrm{matter}}\right] \nonumber \\ \nonumber \\
&& +\Box \delta
R+\left\{\left[\frac{F'''(R)}{F''(R)}-\left(\frac{F'''(R)}{F''(R)}\right)^2\right]
g^{\mu\nu}\nabla_{\mu}R\nabla_{\nu}R
+\frac{R}{3}-\frac{F'(R)}{3F''(R)}
\right. \nonumber \\
&& \left. +\frac{F'''(R)}{3(F''(R))^2}\left[ 2F(R)-RF'(R)+\kappa^2 T^{\mathrm{matter}}
\right] -\frac{\kappa^2}{3F''(R)}\frac{d T^{\mathrm{matter}}}{d R}\right\}\delta R \nonumber\\
& & +2\frac{F'''(R)}{F''(R)}g^{\mu\nu}\nabla_{\mu}R\nabla_{\nu}\delta
R +\mathcal{O}(\delta R^2)\simeq 0\,.\label{completeEq}
\end{eqnarray}
This equation can be used to study perturbations around critical
points.
By assuming $R=R_{0}\simeq \mathrm{const}$ (local approximation), and
$\delta R/R_{0} \ll 1$, we get
\begin{equation}
\Box \delta R\simeq m^2\delta R+\mathcal{O}(\delta R^2)\,,
\label{boxdeltaR}
\end{equation}
where
\begin{equation}
m^2=\frac{1}{3}\left[\frac{F'(R_{0})}{F''(R_{0})}-R_{0}
+\frac{\kappa^2}{F''(R_{0})}
\frac{d T^{\mathrm{matter}}}{d R}\Big\vert_{R_{0}}\right]\,.\label{msquare}
\end{equation}
Here, Eq.~(\ref{criticalpoint2}) with $R_{\mathrm{CP}}=R_{0}$ has
been used.
Note that
\begin{equation}
m^2=\frac{\partial^2 V_{\mathrm{eff}}}{\partial
F'(R)^2}\Big\vert_{R_{0}}\,.
\end{equation}
The second derivative of the effective potential represents the
effective mass of the scalaron. Thus, if $m^2>0$ (in the sense of the quantum
theory, the scalaron, which is a new scalar degree of freedom, is not a
tachyon), one gets a stable solution. For the case of the de Sitter solution,
$m^2$ is positive provided
\begin{equation}
\frac{F'(R_{\mathrm{dS}})}{R_{\mathrm{dS}}
F''(R_{\mathrm{dS}})}>1\,.\label{dSstability}
\end{equation}
The value of the above mass will be later used to check the emergence
of the Newtonian regime.

\subsection{Modified FRW dynamics}

The spatially-flat Friedman-Robertson-Walker (FRW) space-time is
described by the metric
\begin{equation}
ds^2=-dt^2+a^{2}(t)d {\bf x}^2\,,\label{FRWmetric}
\end{equation}
where $a(t)$ is the scale factor of the universe. The Ricci scalar is
\begin{equation}
R=6 \left(2H^2+\dot{H}\right)\,.\label{R}
\end{equation}
In the FRW background, from $(\mu,\nu)=(0,0)$ and the trace part of
the $(\mu,\nu)=(i,j)$ $(i,j=1,...,3)$ components in
Eq.~(\ref{EqEinstMod}), we obtain the equations of motion:
\begin{equation}
\rho_{\mathrm{eff}}=\frac{3}{\kappa^2}H^2\,,\label{FRW1}
\end{equation}
\begin{equation}
p_{\mathrm{eff}}=-\frac{1}{\kappa^2}\left(2\dot{H}+3H^2\right)\,,\label{FRW2}
\end{equation}
where $\rho_{\mathrm{eff}}$ and $p_{\mathrm{eff}}$ are the total effective energy
density and pressure of matter and geometry, respectively, given by
\begin{equation}
\rho_{\mathrm{eff}}=\frac{1}{F'(R)}\left\{\rho+\frac{1}{2\kappa^2}
\left[(F'(R)R-F(R))-6H\dot{F}'(R)\right]\right\}\,,\label{rho}
\end{equation}
\begin{equation}
p_{\mathrm{eff}}=\frac{1}{F'(R)}\left\{p+\frac{1}{2\kappa^2}\left[-(F'(R)R-F
(R))+4H\dot{F}'(R)
+2\ddot{F}'(R)\right]\right\}\,.\label{p}
\end{equation}
Here, $H=\dot{a}(t)/a(t)$ is the Hubble parameter and the dot denotes time
derivative $\partial/\partial t$. This is the form that the total stress tensor
in Eq.~(\ref{EqEinstMod}) assumes in the FRW space-time.

The standard matter conservation law is
\begin{equation}
\dot{\rho}+3H(\rho+p)=0\,,
\end{equation}
and, for a perfect fluid,
\begin{equation}
p=\omega\rho\,,
\end{equation}
$\omega$ being the thermodynamical EoS-parameter of matter.


We also introduce the effective EoS by using the corresponding parameter
$\omega_{\mathrm{eff}}$
\begin{equation}
\omega_{\mathrm{eff}}=\frac{p_{\mathrm{eff}}}{\rho_{\mathrm{eff}}}\,,
\end{equation}
and get
\begin{equation}
\omega_{\mathrm{eff}}=-1-\frac{2\dot{H}}{3H^2}\,.\label{omegaeff}
\end{equation}
If the strong energy condition (SEC) is satisfied
($\omega_{\mathrm{eff}}>-1/3$),
the universe expands in a decelerated way, and vice-versa.
We are interested in the accelerating FRW cosmology below.

\subsection{Spherically symmetric solutions}

In this section we investigate spherically-symmetric solutions (like the Schwarschild black hole),
which constitute an essential element for the local tests of modified gravity under consideration.
For the metric, we start from a static, spherically symmetric ansatz of the type,
\begin{equation}
ds^2=-C(r)\mathrm{e}^{2\alpha(r)}dt^2+\frac{dr^2}{C(r)}+r^2d\Omega\,,
\end{equation}
where $d\Omega=r^2(d\theta^2+\sin^2\theta d\phi^2)$ and $\alpha(r)$ and
$C(r)$ are functions of $r$.

Plugging this ansatz into the action (\ref{action}), and noting that
\begin{eqnarray}
R_{\mathrm{sp}} &=& -3\,\left[{\frac{d}{dr}}C\left(r\right)\right]{\frac{d}{dr}}
\alpha\left(r\right)-2\,C\left(r\right)\left[{\frac{d}{dr}}
\alpha\left(r\right)\right]^{2}-{\frac{d^{2}}{d{r}^{2}}}
C(r)-2\,C\left(r\right){\frac{d^{2}}{d{r}^{2}}}\alpha\left(r\right)\nonumber\\
&&-\frac{4}{r}\,{\frac{d}{dr}}C\left(r\right)
 -\frac{4\,C\left(r\right)}{r}\,{\frac{d}{dr}}\alpha\left(r\right)
 -{\frac{2\, C\left(r\right)}{{r}^{2}}}
+\frac{2}{{r}^{2}}\,,\label{tre}
\end{eqnarray}
one arrives at the  following equations of motion (in vacuum) \cite{seba}:
\begin{eqnarray}
&&\frac{R_\mathrm{sp}F'(R_\mathrm{sp})-F(R_\mathrm{sp})}{F'(R_\mathrm{sp})}
 -2\frac{\left(1-C(r)-r(dC(r)/dr)\right)}{r^2} \\
&& +\frac{2C(r)F''(R_\mathrm{sp})}{F'(R_\mathrm{sp})}\left[
\frac{d^2 R_\mathrm{sp}}{d r^2}+\left(\frac{2}{r}+\frac{dC(r)/dr}{2 C(r)}\right)
\frac{d R _\mathrm{sp}}{d r}+\frac{F'''(R_\mathrm{sp})}{F''(R_\mathrm{sp})}
\left(\frac{d R_\mathrm{sp}}{d r}\right)^2\right]=0\,, \nonumber
\label{one}
\end{eqnarray}
\begin{equation}
\left[\frac{d\alpha(r)}{dr}\left(\frac{2}{r}+\frac{F''(R_\mathrm{sp})}{F'(R_\mathrm{sp})}
\frac{dR_\mathrm{sp}}{d r}\right)-\frac{F''(R_\mathrm{sp})}{F'(R_\mathrm{sp})}
\frac{d^2 R_\mathrm{sp}}{d r^2}-\frac{F'''(R_\mathrm{sp})}{F'(R_\mathrm{sp})}\left(
\frac{d R_\mathrm{sp}}{d r}\right)^2\right]=0\,.\label{two}
\end{equation}
These equations form a system of ordinary differential equations in the three unknown
quantities $\alpha(r)$, $C(r)$, and $R_\mathrm{sp}(r)$.
When $F(R)=R$, the above  system of differential equations lead to
the Schwarzschild solution, namely
\begin{equation}
\alpha(r)=\mathrm{const}\,,\label{S1}
\end{equation}
\begin{equation}
C(r)=\left(1-\frac{2M}{r}\right)\,,\label{S2}
\end{equation}
with $M$ a dimensional constant, and $R_{\mathrm{sp}}=0$.

Another well known case is the one associated with $R_\mathrm{sp}$ being constant. As a result, with  $\alpha=0$, Eq.~(\ref{two}) is trivially satisfied, and the other two equations lead to the Schwarschild-de Sitter solution
\begin{equation}
C(r)=\left(1-\frac{2M}{r}-\frac{\Lambda
r^2}{3}\right)\,,\label{S2bis}
\end{equation}
with
\begin{equation}
R_{\mathrm{sp}}=4\Lambda\,,\label{quattro}
\end{equation}
and
\begin{equation}
2 F(R_{\mathrm{sp}})=R_{\mathrm{sp}}
F'(R_{\mathrm{sp}})\,.\label{zurp}
\end{equation}

\setcounter{equation}{0}
\section{Viability conditions in $F(R)$-gravity \label{SectIII}}

The viability conditions follow from the fact that
the theory is consistent with the results of General Relativity if
$F(0)=0$.
In this way we can have the Minkowski solution.
Recall that in order to avoid anti-gravity effects, it is required that
$F'(R)>0$, namely, the positivity of the effective gravitational coupling.

\subsection{Existence of a matter era and stability of cosmological
perturbations}

On the critical points, $\dot{F'}(R)=0$, and from Eqs.~(\ref{rho}) and (\ref{p}), one has
\begin{equation}
\rho_{\mathrm{eff}}=\frac{1}{F'(R)}\left\{\rho+\frac{1}{2\kappa^2}
\left[(F'(R)R-F(R))\right]\right\}\,,\label{rho1}
\end{equation}
\begin{equation}
p_{\mathrm{eff}}=\frac{1}{F'(R)}\left\{p+\frac{1}{2\kappa^2}\left[-(F'(R)R-F
(R))\right]\right\}\,.\label{p1}
\end{equation}
During the matter dominance era, we have  $p_{\mathrm{eff}}=0$. As as consequence, neglecting the contribution
of the radiation, namely $p=0$, one has $\rho_{\mathrm{eff}}=\rho/F'(R) $ and
\begin{equation}
\frac{RF'(R)}{F(R)}=1\,,\label{M1}
\end{equation}
thus
\begin{equation}
\frac{d}{d R}\left(\frac{RF'(R)}{F(R)}\right)=0\,,
\end{equation}
and using Eq.~(\ref{M1}), this leads to
\begin{equation}
\frac{F''(R)}{F'(R)}=0\,,
\end{equation}
so that, during the matter era, we have $F''(R)\simeq 0$.

In order to reproduce the results of the standard model, where
$R=\kappa^2\rho$ when matter drives the cosmological expansion, a
$F(R)$-theory is acceptable if the modified gravity contribution
vanishes during this era and $F'(R)\simeq 1$.
However, another condition is required on the second derivative of
$F(R)$: it not only has to be very small, but also positive. This last
condition arises from the stability of the cosmological perturbations. We
consider a small region of space-time in the weak-field regime, so that the
curvature is approximated by $R\simeq R_0+\delta R$,
where $R_0\simeq \mathrm{const}$. From Eq.~(\ref{criticalpoint2}), we
obtain
\begin{equation}
-\kappa^2 T=F'(R)R+2(F(R)-RF'(R))\,
\end{equation}
and, since $F'(R)\simeq 1$ and $|(F(R)-R)| \ll R$, we can expand this
expression as
\begin{equation}
-\kappa^2 T\simeq = -\kappa^2 \left(T|_{R} + dT \right) =
R+2(F'(R)-1)\delta R\,,
\end{equation}
with $(\delta R/R) \ll 1$. By evaluating it at $R=R_{0}$,
from Eq.~(\ref{msquare}) one has
\begin{equation}
m^2\simeq
\frac{1}{3}\left(\frac{2}{F''(R_{0})}-\frac{F'(R_{0})}{F''(R_{0})}-R_{0}\right)
\simeq\frac{1}{3}\left(\frac{1}{F''(R_{0})}-\frac{(F'(R_{0})-1)}{F''(R_{0})}
\right)\,,
\end{equation}
which is in agreement with Ref.~\cite{Faraoni}. Since $(F'(R_{0})-1)
\ll 1$ and as $F''(R_0)$ is very close to zero, if $F''(R_0)<0$ the theory will be
strongly unstable. Thus, we have to require $F''(R)>0$ during the matter era.

\subsection{Existence and stability of a late-time de Sitter point}

It is convenient to introduce the following function, $G(R)$,
\begin{equation}
G(R)=2F(R)-RF'(R)\,. \label{G(R)}
\end{equation}
On the zeros of $G(R)$ we recover the condition in Eq.~(\ref{dScondition})
and we have the de Sitter solution which describes the accelerated expansion of
our universe. If the condition in Eq.~(\ref{dSstability}) is satisfied, the
solution will be stable.

A reasonable theory of modified gravity which reproduces the current
acceleration of the universe needs to lead to an accelerating solution
for $R=4\Lambda$, $\Lambda$ being the cosmological constant (typically
$\Lambda\simeq 10^{-66}\text{eV}^2$).
Recall that in the  de Sitter case,  the EoS parameter
$\omega_{\mathrm{eff}}=\omega_\mathrm{dS}=-1$, and all available
cosmological data confirm that its value is actually very close to $-1$. The
possibility of an effective quintessence/phantom dark energy is not
excluded, but the most realistic solution for our current universe is
a (asymptotically) stable de Sitter solution.

\subsection{Local tests and the stability on a planet's surface}

GR was first confirmed by accurate local tests
at the level of the Solar System. A theory of modified gravity has to
admit an asymptotically flat (this is important in order to define the mass
term) static spherically-symmetric solution of the type
(\ref{S2bis}), with $\Lambda$ very small. The typical value of the curvature in
the Solar System, far from sources, is $R=R^{*}$, where $R^*\simeq 10^{-61}
\text{eV}^2$ (it corresponds to one hydrogen atom per cubic centimeter). If a
Schwarzshild-de Sitter solution exists, it will be stable provided
\begin{equation}
\frac{F'(R^{*})}{R^*F''(R^{*})}>1\,.\label{bip}
\end{equation}
The stability of the solution is necessary in order to find the
post-Newtonian parameters in GR.

Concerning the matter instability \cite{Faraoni,Nojiri2003,matterinstability},
this might also occur when the curvature is rather large, as on a planet
($R\simeq 10^{-38} \text{eV}^2$), as compared with the average curvature of the
universe today ($ R\simeq 10^{-66} \text{eV}^2$). In order to arrive to a
stability condition, we
can start from Eq.~(\ref{completeEq}), where $R=R_{b}$ assumes the
typical
curvature value on the planet and $\delta R$ is a perturbation due to
the
curvature difference between the internal and the external solution.
Since
$R_b\simeq -\kappa^2 T^{\mathrm{matter}}$ and $\delta R$ depends on time only, one has
\begin{equation}
-\partial_{t}(\delta R)\simeq \mathrm{const} + U(R_{b})\delta R\,,
\end{equation}
where
\begin{eqnarray}
& &
U(R_{b})=\left\{\left[\left(\frac{F'''(R_{b})}{F''(R_{b})}\right)^2
-\frac{F'''(R_{b})}{F''(R_{b})}\right]g^{\mu\nu}\nabla_{\mu}R_{b}\nabla_{\nu}R_{b}
-\frac{R_{b}}{3}+\frac{F'(R_{b})}{3F''(R_{b})}\right.\nonumber \\
& &
-\left. \frac{F'''(R_{b})}{3(F''(R_{b}))^2}(2F(R_{b})-R_{b}F'(R_{b})-R_{b})\right\}
\delta
R\,.\label{U}
\end{eqnarray}
If $U(R_{b})$ is negative, then the perturbation $\delta R$ becomes
exponentially large and the whole system becomes unstable. Thus, the
matter stability condition is
\begin{equation}
U(R_{b})>0,\phantom{space}\text{ where }\phantom{space}R_{b}\simeq
10^{-38}
\text{eV}^2\,.
\end{equation}
At the cosmological level this means that $F''(R)\simeq 0^+$, in the matter era.
If $F(R\simeq R_{b})\simeq R$, Eq.~(\ref{U}) reads, simply,
$U(R_{b})=1/(3F''(R_{b}))$.

\subsection{Existence of an early-time acceleration and the future
singularity problem}

In order to reproduce the early-time acceleration of our universe,
namely
the inflation epoch, the modified gravity models have to admit a
solution for $\omega_{\mathrm{eff}}$
in Eq.~(\ref{omegaeff}) smaller than $-1/3$. An important point is
that this solution should be unstable.

If the model reproduces the de Sitter solution when $R=R_{\mathrm{dS}}\simeq
10^{20-38} \text{GeV}^2$ (this is the typical curvature value at
inflation),
we have to require that Eq.~(\ref{dSstability}) is violated. Thus,
the
characteristic time of the instability $t_i$ is given by the inverse
of the mass of the scalaron in Eq.~(\ref{msquare}):
\begin{equation}
t_{i}\simeq \Big\vert\frac{1}{m}\Big\vert
=\Big\vert\sqrt{\frac{F'(R_{\mathrm{dS}})}{F'(R_{\mathrm{dS}})-R_{\mathrm{dS
}}
F''(R_{\mathrm{dS}})}}\Big\vert\,.\label{ti}
\end{equation}
Note that if the scalaron mass is equal to zero,
a more detailed analysis, as in Sect.~\ref{detailed}, is needed.

Furthermore, it is well-know that many of the effective quintessence/phantom
dark energy models, including modified gravity, bring the future
universe evolution to a finite-time singularity.
The most familiar of them is the famous Big Rip \cite{Rip}, which is caused by
phantom dark energy. Finite-time future singularities in modified gravity have
been studied in Refs.~\cite{singF(R),singF(G)}. As is known,
a finite-time future singularity occurs when some physical quantity
(as, for instance, the scale factor, effective energy-density or pressure of the
universe or, more simply, some of the components of the Riemann tensor) diverges.
The classification of the (four) finite-time future singularities has been
done in Ref.~\cite{classificationSingularities}. Some of these future
singularities are softer than others and not all physical quantities necessarily
diverge on the singularity.

The presence of a finite-time future singularity may cause serious
problems to the cosmological evolution or to the corresponding black hole or stellar
astrophysics \cite{Maeda}. Thus, it is always necessary to avoid such scenario in realistic
models of modified gravity. It is remarkable that modified gravity actually provides
a very natural way to cure such singularities by adding, for instance, an
$R^2$-term \cite{abdalla,singF(R)}. Simultaneously with the removal of any possible
future singularity, the addition of this term supports the early-time inflation
caused by modified gravity. Remarkably, even in the case inflation were not an element
of the alternative gravity dark energy model considered, it will eventually occur after
adding such higher-derivative term. Hence, the removal of future singularities is
a natural prescription for the unified description of the inflationary and dark energy epochs.

\setcounter{equation}{0}
\section{Realistic exponential gravity \label{SectIV}}

In Refs.~\cite{Sawiki,Battye,MioUno} several versions of viable
modified
gravity have been proposed, the so-called one-step models, which
reproduce the current acceleration of the universe. They incorporate
a vanishing (or fast decreasing) cosmological constant in the flat
($R\rightarrow 0$) limit, and exhibit a suitable, constant asymptotic
behavior for large values of $R$. The simplest one was proposed in
Ref.~\cite{MioUno}
\begin{equation}
F(R)=R-2\Lambda \left(1-\mathrm{e}^{-R/R_0}\right)\,.\label{model}
\end{equation}
Here, $\Lambda\simeq 10^{-66}\text{eV}^2$ is the cosmological
constant and
$R_{0}\simeq \Lambda$ a curvature parameter. In flat space one has
$F(0)=0$
and recovers the Minkowski solution. For $R \gg R_{0}$, $F(R)\simeq
R-2\Lambda$, and the theory mimics the $\Lambda$CDM model. Note that
late-time cosmology of such exponential gravity was also considered
in Ref.~\cite{Linder}.

For simplicity, we will set
\begin{equation}
f(R)=-2\Lambda(1-\mathrm{e}^{-R/R_0})\,,
\end{equation}
and thus
\begin{equation}
f'(R)=-2\frac{\Lambda}{R_{0}}\mathrm{e}^{-R/R_0}\,,
\end{equation}
\begin{equation}
f''(R)=2\frac{\Lambda}{R_{0}^2}\mathrm{e}^{-R/R_0}\,.
\end{equation}
Since $|f'(R \gg R_{0})| \ll 1$, the model is protected against
anti-gravity during the whole
cosmological evolution, until the de Sitter solution ($R= 4\Lambda$)
of today's universe is reached.
For large values of the curvature, $F(R \gg R_{0})\simeq R$, and we
can reconstruct the matter-dominated era, as in GR. In particular, $F''(R
\gg R_{0})\sim \Lambda \mathrm{e}^{-R/R_0}\simeq 0^+$, and we do not have any
instability problems related to the matter epoch, obtaining matter stability
on a planet's surface and at the solar system scale.

Let us consider the $G(R)$ function of Eq.~(\ref{G(R)}),
\begin{equation}
G(R)=R+2f(R)-Rf'(R)\,.
\end{equation}
Since $G(0)=0$, one has a trivial de Sitter solution for $R=0$.
Consider now
\begin{equation}
G'(R)=1+f'(R)-Rf''(R)\,.
\end{equation}
Since $G'(0)<0$, the function $G(R)$ becomes negative and starts to
increase
after $R= R_{0}$.
For $R=4\Lambda$, $F(R)\simeq -2\Lambda$, $F'(R)\simeq 1$ and
$F''(R)\simeq
0^+$.
It means that $G(4\Lambda)\simeq 0$ and we find the de Sitter
solution of
the dark energy phase which is able to describe the current
acceleration of
our universe. After this stage, $G(R>4\Lambda)\simeq R>0$ and we do
not find
other de Sitter solutions. Note that the de Sitter solution for
$R=4\Lambda$
is stable, since the first term in Eq.~(\ref{dSstability}) diverges.
On the other hand, the Minkowski space solution is unstable. Summing
up,
we have two FRW-vacuum solutions, which correspond to the trivial de
Sitter
point for $R=0$ and to the stable de Sitter point of current
acceleration,
for $R=4\Lambda$.

Finally, we have to consider the existence of spherically-symmetric
solution. In $R=0$ we find the Schwarzschild solution, which is
unstable. On
the other hand, the physical Schwarzschild-de Sitter solutions are
obtained
for $R \gg R_{0}$. For example, in the Solar System, $R^*\simeq
10^{-61}
\text{eV}^2$.
In this case ($F(R^*)\simeq R^*-2\Lambda$) we find the
Schwarzschild-de
Sitter solution
as in Eq.(\ref{S2bis}),
which can be approximated with the Schwarzschild solution of
Eqs.~(\ref{S1})-(\ref{S2}),
owing to the fact that $\Lambda$ is very small. For $R^* \gg R_{0}$,
Eq.~(\ref{bip}) with $R_{\mathrm{sp}}=R^*$ is satisfied and the
solution is stable.

The description of the cosmological evolution in exponential gravity
has
been carefully studied in Refs.~\cite{Linder,Bamba} where it was
explicitly demonstrated that the late-time cosmic acceleration
following the
matter-dominated stage, as final attractor of the universe, can
indeed be
realized. By carefully fitting the value of $R_{0}$, the correct
value of
the rate between matter and dark energy of the current universe
follows (see Sect.~\ref{SectVI}).
Our next step will be to generalize the model in order to describe inflation.
We will follow the method first suggested in Ref.~\cite{MioUno}.

\setcounter{equation}{0}
\section{Inflation \label{SectV}}

A simple modification of the one-step model which incorporates the
inflationary era is given by a combination of the function discussed
above
with another one-step function reproducing the cosmological constant
during inflation. A quite natural possibility is
\begin{equation}
F(R)=R-2\Lambda\left(1-\mathrm{e}^{-\frac{R}{R_{0}}}\right)
-\Lambda_{i}\left(1-\mathrm{e}^{-\left(\frac{R}{R_i}\right)^n}\right)
+\gamma R^\alpha\,. \label{total}
\end{equation}
For simplicity, we call
\begin{equation}
f_{i}=
-\Lambda_{i}\left(1-\mathrm{e}^{-\left(\frac{R}{R_i}\right)^n}\right)\,,
\end{equation}
where $R_{i}$ and $\Lambda_{i}$ assume the typical values of the
curvature
and expected cosmological constant during inflation, namely $R_{i}$,
$\Lambda_i$ $\simeq 10^{20-38} \text{eV}^2$, while $n$ is a natural
number
larger than one. The presence of this additional parameter is
motivated by the necessity to avoid the effects
of inflation during the matter era, when $R \ll R_{i}$, so that, for
$n>1$,
one gets
\begin{equation}
R \gg |f_i(R)|\simeq \frac{R^{n}}{R_i^{n-1}} \,.
\end{equation}
The last term in Eq.~(\ref{total}), namely $\gamma R^\alpha$, where
$\gamma$
is a positive dimensional constant and $\alpha$ a real number, is
necessary
to obtain the exit from inflation. If $\gamma\sim 1/R_i^{\alpha-1}$
and
$\alpha>1$,
the effects of this term vanish in the small curvature regime, when
$R \ll
R_i$ and
\begin{equation}
R \gg \frac{R^\alpha}{R_i^{\alpha-1}}\,.
\end{equation}
Note that $f_{i}(0)=0$ and $f_{i}(R \gg R_i)\simeq -\Lambda_i$. We
also
obtain
\begin{equation}
f_i'(R)=-\frac{\Lambda_i n
R^{n-1}}{R_i^n}\mathrm{e}^{-\left(\frac{R}{R_i}\right)^n}\,,
\end{equation}
\begin{equation}
f_i''(R)=-\frac{\Lambda_i
n(n-1)R^{n-2}}{R_i^n}\mathrm{e}^{-\left(\frac{R}{R_i}\right)^n}
+\Lambda_i\left(\frac{n R^{n-1}}{R_i^n}\right)^2
\mathrm{e}^{-\left(\frac{R}{R_i}\right)^n}\,.\label{secondderivative}
\end{equation}
The first derivative $f_i'(R)$ has a minimum at $R=\tilde R$, where
$f_i''(\tilde R)=0$. One gets
\begin{equation}
\tilde{R}=R_i\left(\frac{n-1}{n}\right)^{\frac{1}{n}}\,.\label{Rstar}
\end{equation}
Thus, in order to avoid the anti-gravity effects ($|f'_i(R)|<1$), it
is sufficient to require $|f'_i(\tilde{R})|<1$. This leads to
\begin{equation}
R_i>\Lambda_i n
\left(\frac{n-1}{n}\right)^{\frac{n-1}{n}}\mathrm{e}^{-\frac{n-1}{n}}\,.\label{uno}
\end{equation}
For example, one can choose $n=4$. In this case Eq.~(\ref{uno}) is
satisfied for $R_i>1.522\, \Lambda_i$.
A reasonable choice is $R_i=2\Lambda_i$.
The last power-term of Eq.~(\ref{total}) does not give any problem
with anti-gravity, because its first derivative is positive.

It is necessary that the modification of gravity describing inflation
does not have any influence on the stability of the matter era in the
small curvature range. When $R \ll R_i$, the second derivative of such
modification, namely
\begin{equation}
f''_i(R)+\alpha(\alpha-1)\gamma R^{\alpha-2}\simeq
\frac{1}{R}\left[-n(n-1)\left(\frac{R}{R_i}\right)^{n-1}
+\alpha(\alpha-1)\left(\frac{R}{R_i}\right)^{\alpha-1}\right]\,,
\label{zwei}
\end{equation}
must be positive, that is
\begin{equation}
n>\alpha\,.
\end{equation}

We require the existence of the de Sitter critical point
$R_{\mathrm{dS}}$
which describes inflation in the high-curvature regime of $f_i(R)$,
so that
$f_i(R_{\mathrm{dS}} \gg R_i)\simeq -\Lambda_i$
and $f_i'(R_{\mathrm{dS}} \gg R_i)\simeq 0^+$. In this region, the
role of
the first term of Eq.~(\ref{total}) is negligible, while the term
$\gamma
R^{\alpha}$ needs to be taken into account. For simplicity, we shall
assume
that $\gamma=1/R_{\mathrm{dS}}^{\alpha-1}$.
The function $G(R)$ in Eq.~(\ref{G(R)}),
\begin{equation}
G(R)=R+2f_i-Rf_i'+\frac{(2-\alpha)}{R_{\mathrm{dS}}^{\alpha-1}}R^\alpha\,,\label{ughino}
\end{equation}
has to be zero on the de Sitter solution. We get
\begin{equation}
R_{\mathrm{dS}}=\frac{2\Lambda_{i}}{3-\alpha}\,,\phantom{space}
\phantom{space}R_{\mathrm{dS}} \gg R_i\,.\label{due}
\end{equation}
Since $\Lambda_i\sim R_i$, in order to satisfy the last two
conditions simultaneously, one has to choose
\begin{equation}
2<\alpha<3\,.
\end{equation}
Let us consider the effective scalar mass of Eq.~(\ref{msquare}) on
the de Sitter solution:
\begin{equation}
m^2\simeq\frac{R_{\mathrm{dS}}}{3}\left(\frac{1+2\alpha-\alpha^2}{\alpha(\alpha-1)}\right)\,.
\end{equation}
It is negative if $\alpha>2.414$. In this case inflation is strongly
unstable. Using Eq.~(\ref{ti}) we derive the characteristic time of
the instability as
\begin{equation}
t_i\sim\frac{1}{\sqrt{R_{\mathrm{dS}}}}\sim 10^{-10}-10^{-19}\, \text{sec}
\,,
\end{equation}
in accordance with the expected value. The new condition on $\alpha$,
in order to have unstable inflation, is
\begin{equation}
5/2\leq\alpha<3\,.
\end{equation}

Now, we will try to reconstruct the evolution of the function $G(R)$
in Eq.~(\ref{ughino}),
\begin{equation}
G(R)=R+2f_i(R)-Rf'_{i}(R)+(2-\alpha)\frac{R^\alpha}{R_{\mathrm{dS}}^{\alpha-
1}}\,.
\end{equation}
When $R=0$, we find a trivial de Sitter point and $G(0)=0$.
For the first derivative of $G(R)$,
\begin{equation}
G'(R)=1+f'_i(R)-Rf''_i(R)
+\alpha(2-\alpha)\frac{R^{\alpha-1}}{R_{\mathrm{dS}}^{\alpha-1}}\,.
\label{ugo}
\end{equation}
$G'(0)>0$ and $G(R)$ increases. Since $f_i''(R)$ starts being
positive for
$R>\tilde{R}$
(where $\tilde{R}$ is expressed as in Eq.~(\ref{Rstar})) and
$2-\alpha<0$,
it is easy to see that $G(R)$ begins to decrease at around $R=R_i$
and that
it is zero when $R=R_{\mathrm{dS}}$.
After this point, $G'(R>R_{\mathrm{dS}})<0$ and we do not have other
de
Sitter solutions.
On the other hand, it is possible to have a fluctuation of $G(R)$
along the
$R$-axis just before the de Sitter point describing inflation takes
over. In
order to avoid other de Sitter solutions (i.e., possible final
attractors
for the system), we need to verify the fulfillment of the following
condition:
\begin{equation}
G(R)>0\phantom{space}\text{for}\phantom{space}0<R<R_{\mathrm{dS}}\,.
\end{equation}
Precise analysis of this condition leads to a transcendental
equation.
In the next subsection we will limit ourselves to a graphical
evaluation.
In general, it will be enough to choose $n$ sufficiently large in
order to avoid such effects.

\subsection{Construction of a realistic model for inflation}

By taking into account all the conditions met in the previous
paragraph,
the simplest choice of parameters to introduce in the function of
Eq.~(\ref{total}) is
\begin{equation}
n=4\,,\phantom{space}\alpha=\frac{5}{2}\,,
\end{equation}
while the curvature $R_i$ is set as
\begin{equation}
R_i=2\Lambda_i\,.
\end{equation}
In this way, $n>\alpha$ and we avoid undesirable instability effects
in the small-curvature regime.
$R_i$ satisfies Eq.~(\ref{uno}) and we have no anti-gravity effects.
From Eq.~(\ref{due})
one recovers the unstable de Sitter solution describing inflation as
\begin{equation}
R_{\mathrm{dS}}=4\Lambda_i\,.\label{mitiko}
\end{equation}
We note that, due to the large value of $n$, $R_{\mathrm{dS}}$ is
sufficiently large with respect to $R_i$, and
$f_i(R_{\mathrm{dS}})\simeq
-\Lambda_i$. One can also expect that, on top of this graceful exit from inflation, the
effective scalar degree of freedom may also give rise to reheating, in analogy with
Ref.~\cite{dima}.

\begin{figure}[-!h]
\begin{center}
\includegraphics[angle=0, width=0.5\textwidth]{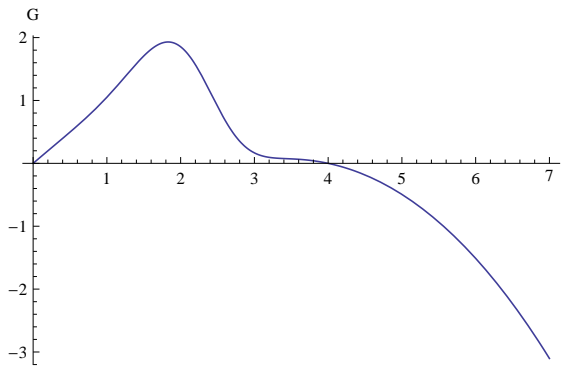}
\caption{Plot of $G(R/\Lambda_i)$
The zeros correspond to the de Sitter solutions. \label{Fig1}}
\end{center}
\end{figure}

In Fig.~\ref{Fig1} a plot of $G(R)$ is shown. The zeros of $G(R)$
correspond to de Sitter solutions. One can see that the only non-trivial zero is
the de Sitter point of Eq.~(\ref{mitiko}), and here the function crosses the
$R$-axis up-down, according to the instability of such solution
(since  $F''(R>\tilde{R})>0$, we get $G'(R)\sim m^2$<0).
This means that the inflationary de Sitter point corresponds to a
maximum of the theory (without matter/radiation).
The system gives rise to the de Sitter solution where the universe
expands in an accelerating way but, suddenly, it exits from inflation and tends towards
the minimal attractor at $R=0$, unless the theory develops a singularity solutions
for $R\rightarrow \infty$. In such case, the model could exit from inflation and move in the
wrong direction, where the curvature would grow up and diverge, and a singularity would appear.
In the next section we will recall some important facts about singularities,
which will be considered in the context of exponential gravity.

\setcounter{equation}{0}
\section{Finite-time future singularities \label{SectVI}}

In general, future singularities appear when the Hubble parameter has
the form
\begin{equation}
H=\frac{h}{(t_{0}-t)^{\beta}}\label{Hsingular}\,,
\end{equation}
where $h$ and $t_{0}$ are positive constants, and $t<t_{0}$, because
it should correspond to an expanding universe. Here $\beta$ is a
positive
constant or a negative non-integer number, so that, when $t$ is close
to $t_{0}$,
$H$ or some derivatives of $H$, and therefore the curvature, become
singular.

Note that such a choice of the Hubble parameter corresponds to an
accelerated
universe, because on the singular solution of Eq.~(\ref{Hsingular})
it is
easy to see that the strong energy condition
($\rho_{\mathrm{eff}}+3p_{\mathrm{eff}}\geq 0$) is always violated
when
$\beta>0$, or for small values of $t$ when $\beta<0$. What means
that, in
any case, a singularity could emerge at the final evolution stage of
an accelerating universe.

The finite-time future singularities can be classified as follows
\cite{classificationSingularities}:
\begin{itemize}
\item Type I (Big Rip): for $t\rightarrow t_{0}$,
$a(t)\rightarrow\infty$,
$\rho_\mathrm{eff}\rightarrow\infty$ and
$|p_\mathrm{{eff}}|\rightarrow\infty$.
It corresponds to $\beta=1$ and $\beta>1$.
\item Type II (sudden):
for $t\rightarrow t_{0}$, $a(t)\rightarrow a_{0}$,
$\rho_\mathrm{{eff}}\rightarrow\rho_{0}$ and $|p_\mathrm{{eff}}|
\rightarrow\infty$.
It corresponds to $-1<\beta<0$.
\item Type III: for $t\rightarrow t_{0}$, $a(t)\rightarrow a_{0}$,
$\rho_\mathrm{{eff}}\rightarrow\infty$ and
$|p_\mathrm{{eff}}|\rightarrow\infty$.
It corresponds to $0<\beta<1$.
\item Type IV: for $t\rightarrow t_{0}$, $a(t)\rightarrow a_{0}$,
$\rho_\mathrm{{eff}}\rightarrow 0$, $|p_\mathrm{{eff}}|
\rightarrow 0$
and higher derivatives of $H$ diverge.
The case in which $\rho$ and/or $p$ tend to finite values is
also included. It corresponds to
$\beta<-1$ but $\beta$ is not any integer number.
\end{itemize}
Here, $a_{0}(\neq 0)$ and $\rho_{0}$ are positive constants.

It is easy to see that in the case of Type I and Type III
singularities we
have $R\rightarrow+\infty$. Those types are the most dangerous in the
cosmological scenario. On the other hand, also the soft Type II
and Type IV ($R\rightarrow \mathrm{const}$) singularities may cause
various problems related , for instance, to the description of stellar
astrophysics.

\subsection{Singularities in exponential gravity}

Let us now consider the exponential model of Eq.~(\ref{total}),
but avoiding the last term, $\gamma R^{\alpha}$.
When $R\rightarrow +\infty$, we have
$F(R\rightarrow +\infty)\simeq R + \mathrm{const}$,
$F'(R\rightarrow +\infty)\simeq 1$, while the high-order derivatives
of $F(R)$
tend to zero in an exponential way. This means that
$\rho_{\mathrm{eff}}$ in
Eq.~(\ref{rho}),
by neglecting the matter contribution, tends to a constant on the
singular solution of
Eq.~(\ref{Hsingular}) with $\beta>0$ (namely, $R\rightarrow+\infty$).
For this reason, neither Type I nor Type III singularities, where
$\rho_{\mathrm{eff}}$ has to diverge, can appear in exponential
gravity.
However, a model of this kind, which mimics the cosmological constant
in the high-curvature regime, can be affected by Type II or IV
singularities.

Regarding Type II singularities, we have to consider the behavior of
the model
when $R$ is negative, which is very different from the case of
positive values of the curvature. This is the reason why this kind of
singularities does not appear:
when $R\rightarrow-\infty$, $\rho_{\mathrm{eff}}$ of Eq.~(\ref{rho})
exponentially diverge, and the sudden singularity, where
$\rho_{\mathrm{eff}}$ has to tend to a constant, cannot be realized.

Also Type IV singularities are not realized: when $R\rightarrow 0^-$,
$F(R\rightarrow 0^-)\simeq R-R/R_{0}$, and it is easy to see that
$\rho_{\mathrm{eff}}$ of Eq.~(\ref{rho}) behaves as
$1/(t_{0}-t)^{\beta+1}$,
and it is larger than $H^2$($= h^2/(t_{0}-1)^{2\beta}$), when
$\beta<-1$.
For this reason, Eq.~(\ref{FRW1})
is inconsistent with Type IV singularities. The argument is valid
also if we
consider the more general case where $H=H_{0}+h/(t_{0}-t)^{\beta}$
and tends
to the positive constant $H_{0}$ in the asymptotic singular limit:
also in this situation $F(R)$ approaches a constant like
$1/(t_{0}-1)^{\beta+1}$, while the time dependent part of $H^2$
behaves as $1/(t_{0}-t)^\beta$.

Consider now adding back the term $\gamma R^\alpha$. This becomes
relevant just
when $R \gg R_i$, so that it can produce Type I or III singularities,
only.
However, in Ref.~\cite{singF(G)} it is explicitly demonstrated that
the model $R+\gamma R^\alpha$ is protected against singularities of
this kind if $2\geq \alpha>1$.

Thus, we have found  our theory to be free from singularities. In
particular, when our model is approximated as
\begin{equation}
F(R)\simeq R-\Lambda_i+\gamma R^{\alpha}\,,
\end{equation}
Type I or III singularities do not occur. When inflation ends,
the model moves to the attractor de Sitter point. In this
way, the small curvature regime arises, the first term of
Eq.~(\ref{total})
becomes dominant and the physics of the $\Lambda$CDM model are
reproduced.

\setcounter{equation}{0}
\section{Dark energy epoch \label{SectVII}}

We will now be interested in the cosmological evolution of the dark
energy density
$\rho_{\mathrm{DE}}=\rho_{\mathrm{eff}}-\rho/F'(R)$ in the case of the two-step model of
Eq.~(\ref{total}),
near the late-time acceleration era describing the current universe. We
assume the spatially-flat FRW metric of Eq.~(\ref{FRWmetric}).
Let us follow the method first suggested in Ref.~\cite{Sawiki} and
more recently used in Ref.~\cite{Bamba}.

To this end, we introduce the variable
\begin{equation}
y_\mathrm{H}\equiv\frac{\rho_{\mathrm{DE}}}{\rho_m^{(0)}}=\frac{H^2}{\tilde{m}^2}-a^{
-3}-\chi
a^{-4}\,.\label{y}
\end{equation}
Here, $\rho_m^{(0)}$ is the energy density of matter at present time,
$\tilde{m}^2$ is the mass scale
\begin{equation}
\tilde{m}^2\equiv\frac{\kappa^2\rho_m^{(0)}}{3}\simeq 1.5 \times
10^{-67}\text{eV}^2\,,
\end{equation}
and $\chi$ is defined as
\begin{equation}
\chi\equiv\frac{\rho_r^{(0)}}{\rho_m^{(0)}}\simeq 3.1 \times
10^{-4}\,,
\end{equation}
where $\rho_r^{(0)}$ is the energy density of radiation at present
(the contribution from radiation is also taken into consideration).

The EoS-parameter $\omega_{\mathrm{DE}}$ for dark energy is
\begin{equation}
\omega_{\mathrm{DE}}=-1-\frac{1}{3}\frac{1}{y_\mathrm{H}}\frac{d y_\mathrm{H}}{d (\ln
a)}\,.\label{eins}
\end{equation}
By combining Eq.~(\ref{FRW1}) with Eq.~(\ref{R}) and using
Eq.~(\ref{y}),
one gets
\begin{equation}
\frac{d^2 y_\mathrm{H}}{d (\ln a)^2}+J_1\frac{d y_\mathrm{H}}{d (\ln a)}+J_2
y_\mathrm{H}+J_3=0\,,\label{superEq}
\end{equation}
where
\begin{equation}
J_1=4+\frac{1}{y_\mathrm{H}+a^{-3}+\chi a^{-4}}\frac{1-F'(R)}{6\tilde{m}^2
F''(R)}\,,
\end{equation}
\begin{equation}
J_2=\frac{1}{y_\mathrm{H}+a^{-3}+\chi a^{-4}}\frac{2-F'(R)}{3\tilde{m}^2
F''(R)}\,,
\end{equation}
\begin{equation}
J_3=-3 a^{-3}-\frac{(1-F'(R))(a^{-3}+2\chi a^{-4})
+(R-F(R))/(3\tilde{m}^2)}{y_\mathrm{H}+a^{-3}+\chi
a^{-4}}\frac{1}{6\tilde{m}^2
F''(R)}\,,
\end{equation}
and thus, we have
\begin{equation}
R=3\tilde{m}^2 \left(\frac{d y_\mathrm{H}}{d \ln
a}+4y_\mathrm{H}+a^{-3}\right)\,.\label{Ricciscalar}
\end{equation}
The parameters of Eq.~(\ref{total}) are chosen as follows:
\begin{equation*}
\Lambda=(7.93)\tilde{m}^2\,,
\end{equation*}
\begin{equation*}
\Lambda_i=10^{100}\Lambda\,,
\end{equation*}
\begin{equation*}
R_i=2\Lambda_i\,,\quad \phantom{sp}n=4\,,
\end{equation*}
\begin{equation*}
\alpha=\frac{5}{2}\,,\quad
\phantom{sp}\gamma=\frac{1}{(4\Lambda_i)^{\alpha-1}}\,,
\end{equation*}
\begin{equation}
R_0=0.6\Lambda\,,\quad 0.8\Lambda\,,\quad \Lambda\,.
\end{equation}
Eq.~(\ref{superEq}) can be solved in a numerical way, in the range of
$R_0\ll R\ll R_i$ (matter era/current acceleration). $y_\mathrm{H}$ is then
found as a function of the red shift $z$,
\begin{equation}
z=\frac{1}{a}-1\,.
\end{equation}
In solving Eq.~(\ref{superEq}) numerically we have taken the following initial
conditions at $z=z_i$
\begin{equation*}
\frac{d y_\mathrm{H}}{d (z)}\Big\vert_{z_i}=0\,,
\end{equation*}
\begin{equation}
y_\mathrm{H}\Big\vert_{z_i}=\frac{\Lambda}{3\tilde{m}^2}\,,
\end{equation}
which correspond to the ones of the $\Lambda$CDM model. This choice
obeys to
the fact that in the high red shift regime the exponential model is
very
close to the $\Lambda$CDM Model. The values of $z_i$ have been chosen
so
that $RF''(z = z_i) \sim 10^{-5}$, assuming $R= 3\tilde{m}^2 (z+1)^3$.
We have $z_i=1.5$, $2.2$, $2.5$ for $R_0=0.6\Lambda$, $0.8\Lambda$,
$\Lambda$, respectively.
In setting the parameters, we have used the last results of the
$W$MAP, BAO and SN surveys \cite{WMAP}.

\begin{figure}[-!h]
\begin{center}
\includegraphics[angle=0, width=0.5\textwidth]{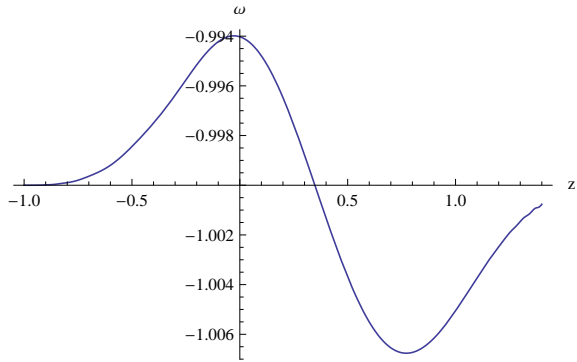}
\end{center}
\caption{Plot of $\omega_\mathrm{DE}$ for $R_0=0.6\Lambda$.
\label{Fig2}}
\end{figure}

\begin{figure}[-!h]
\begin{center}
\includegraphics[angle=0, width=0.5\textwidth]{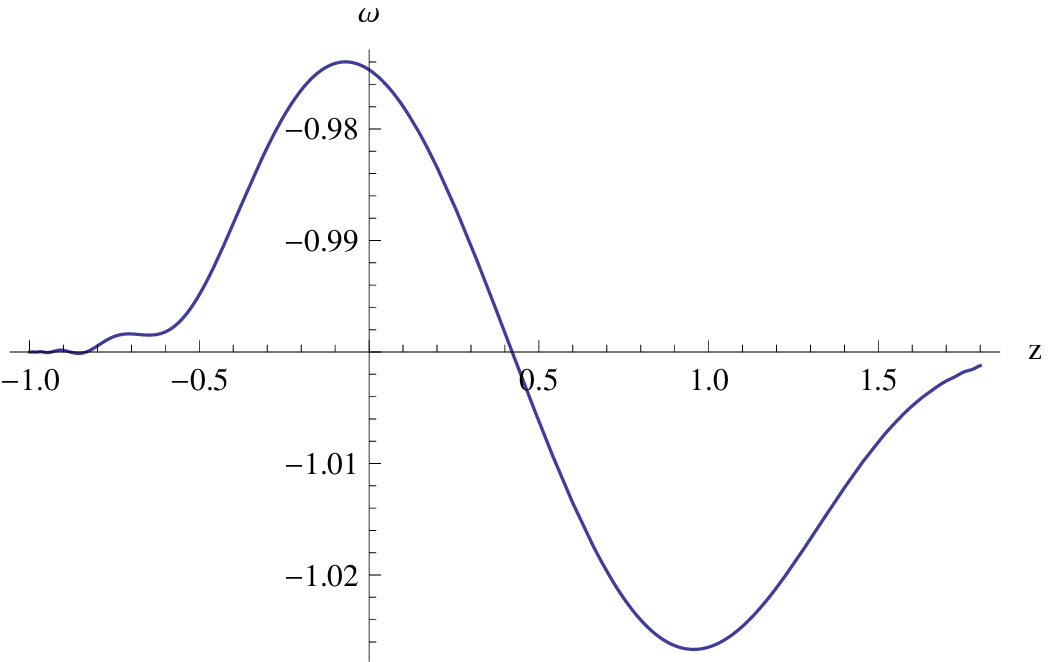}
\end{center}
\caption{Plot of $\omega_\mathrm{DE}$ for $R_0=0.8\Lambda$.
\label{Fig3}}
\end{figure}

\begin{figure}[-!h]
\begin{center}
\includegraphics[angle=0, width=0.5\textwidth]{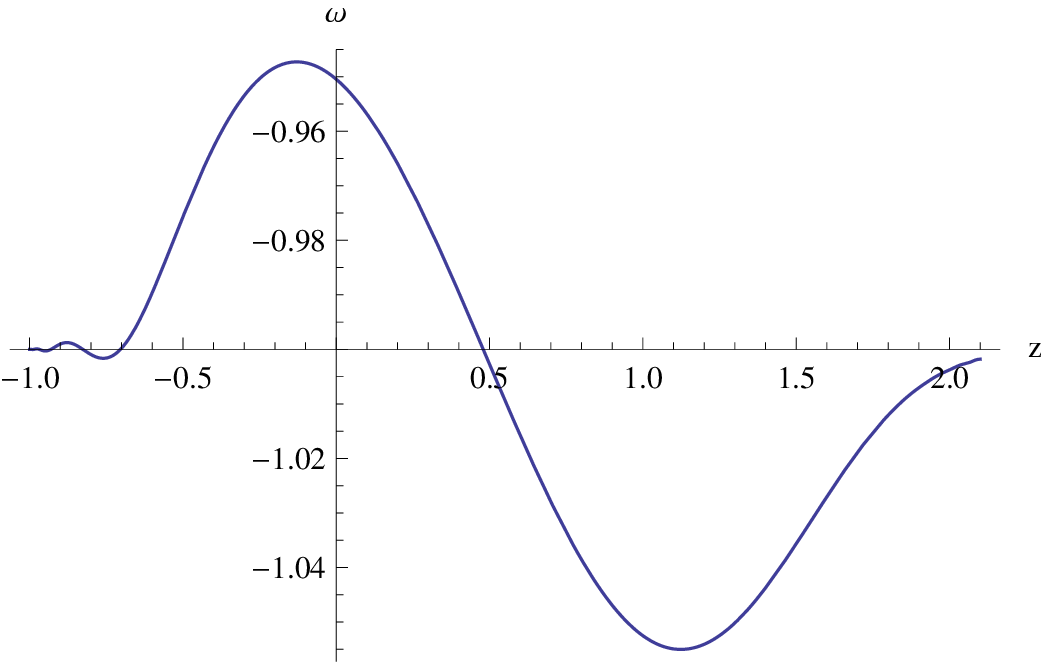}
\end{center}
\caption{Plot of $\omega_\mathrm{DE}$ for $R_0=\Lambda$.
\label{Fig4}}
\end{figure}

Using Eq.~(\ref{eins}), one derives $\omega_\mathrm{DE}$ from
$y_\mathrm{H}$.
In Figs.~\ref{Fig2}, \ref{Fig3} and \ref{Fig4}, we plot
$\omega_\mathrm{DE}$
as a function of the redshift $z$ for
$R_0=0.6\Lambda$, $0.8\Lambda$, $\Lambda$, respectively. Note that
$\omega_\mathrm{DE}$ is very close to minus one. In the present
universe ($z=0$),
one has $\omega_\mathrm{DE}=-0.994$, $-0.975$, $-0.950$ for
$R_0=0.6\Lambda$, $0.8\Lambda$, $\Lambda$. The smaller $R_0$ is, our model
becomes more indistinguishable from the $\Lambda$CDM model, where
$\omega_\mathrm{DE}=-1$.

\begin{figure}[-!h]
\begin{center}
\includegraphics[angle=0, width=0.5\textwidth]{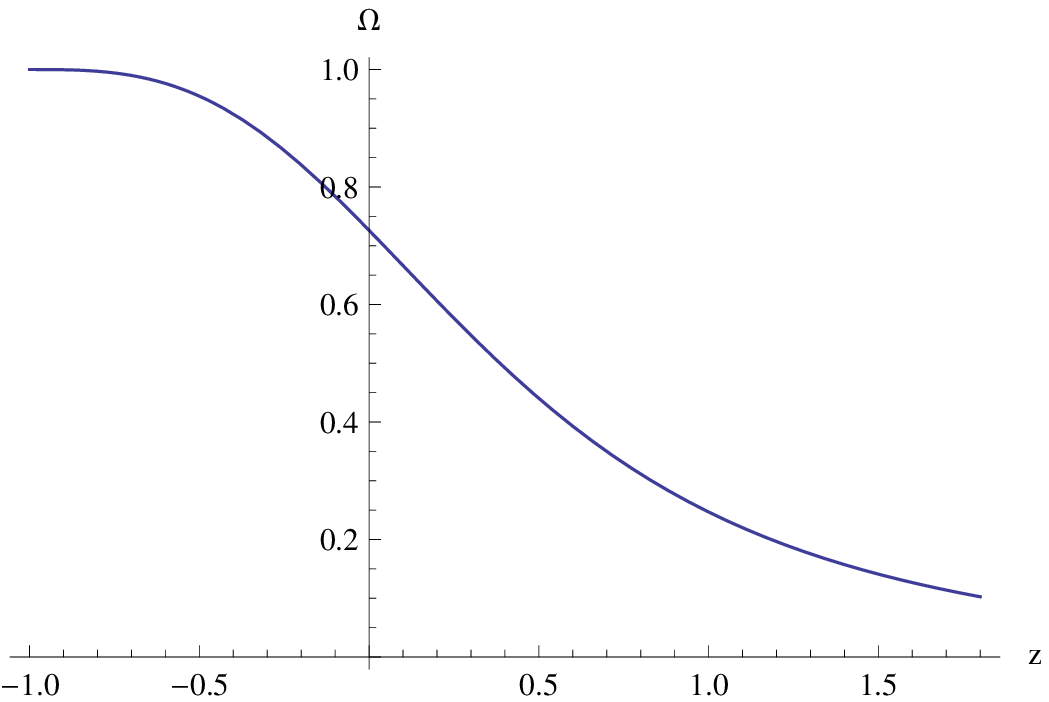}
\end{center}
\caption{Plot of $\Omega_\mathrm{DE}$ for $R_0=0.6\Lambda$.
\label{Fig5}}
\end{figure}

\begin{figure}[-!h]
\begin{center}
\includegraphics[angle=0, width=0.5\textwidth]{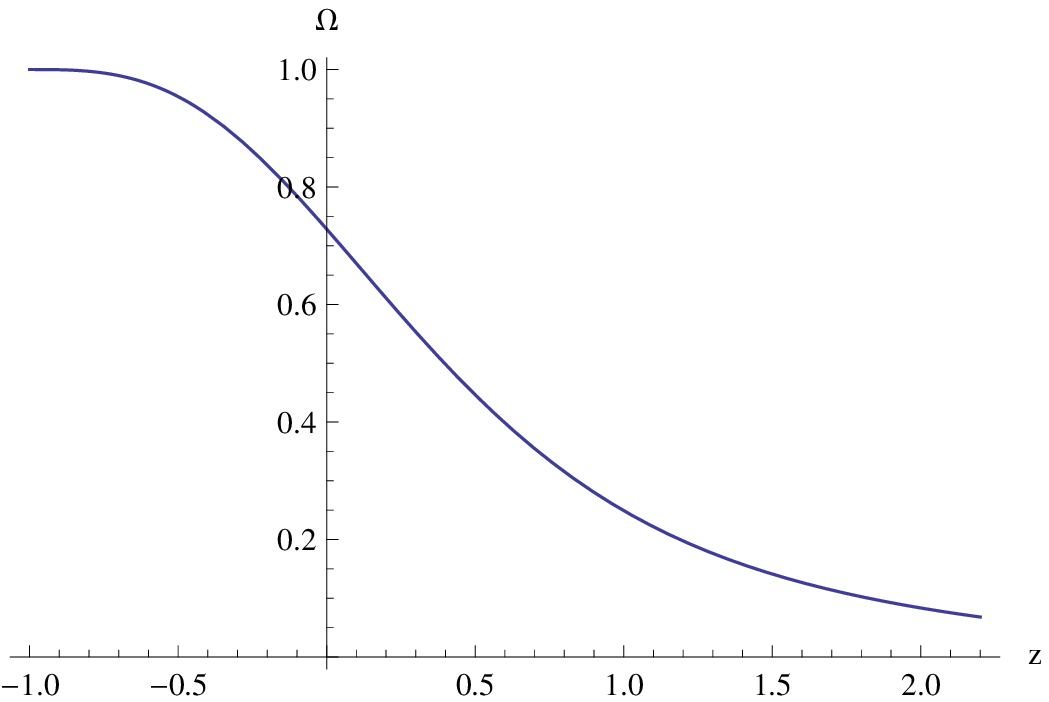}
\end{center}
\caption{Plot of $\Omega_\mathrm{DE}$ for $R_0=0.8\Lambda$.
\label{Fig6}}
\end{figure}

\begin{figure}[-!h]
\begin{center}
\includegraphics[angle=0, width=0.5\textwidth]{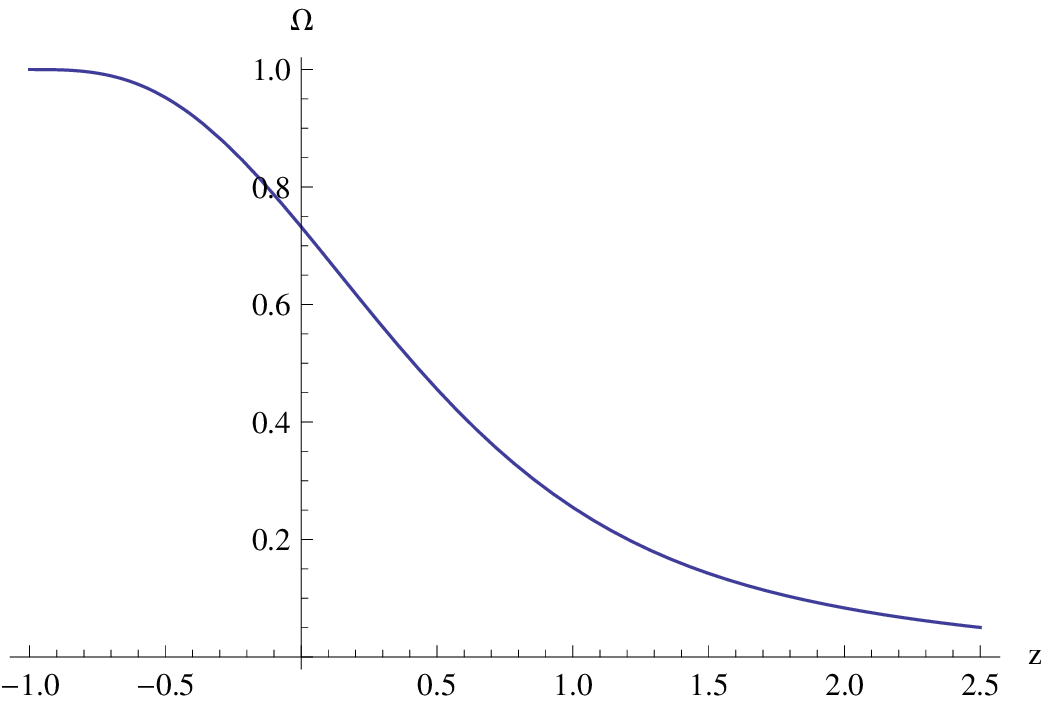}
\end{center}
\caption{Plot of $\Omega_\mathrm{DE}$ for $R_0=\Lambda$.
\label{Fig7}}
\end{figure}

We can also extrapolate the behavior of the density parameter of dark
energy, $\Omega_\mathrm{DE}$,
\begin{equation}
\Omega_\mathrm{DE}\equiv\frac{\rho_\mathrm{DE}}{\rho_\mathrm{eff}}
=\frac{y_\mathrm{H}}{y_\mathrm{H}+\left(z+1\right)^3+\chi\left(z+1\right)^4}\,.
\end{equation}
Plots of $\Omega_\mathrm{DE}$ as a function of
the redshift
$z$ for $R_0=0.6\Lambda$, $0.8\Lambda$, $\Lambda$, are shown in
Figs.~\ref{Fig5},
\ref{Fig6} and \ref{Fig7}. For the present universe ($z=0$), one has
$\Omega_\mathrm{DE}=0.726$, $0.728$,
$0.732$ for $R_0=0.6\Lambda$, $0.8\Lambda$, $\Lambda$, respectively.

The data are in accordance with the last and very accurate
observations of our present universe, where:
\begin{equation*}
\omega_\mathrm{DE}=-0.972^{+0.061}_{-0.060}\,,
\end{equation*}
\begin{equation}
\Omega_\mathrm{DE}=0.721\pm 0.015\,.
\end{equation}

\begin{figure}[-!h]
\begin{center}
\includegraphics[angle=0, width=0.5\textwidth]{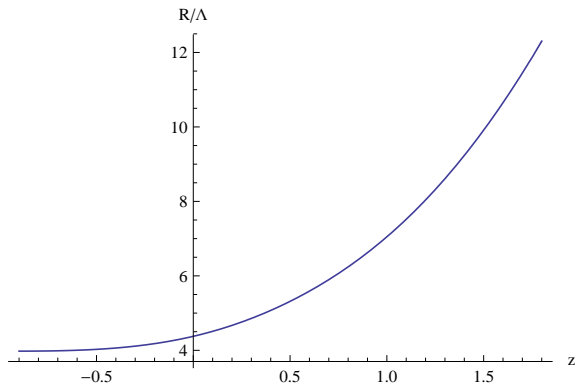}
\end{center}
\caption{Plot of $R/\Lambda$ for $R_0=0.6\Lambda$.
\label{Fig8}}
\end{figure}

\begin{figure}[-!h]
\begin{center}
\includegraphics[angle=0, width=0.5\textwidth]{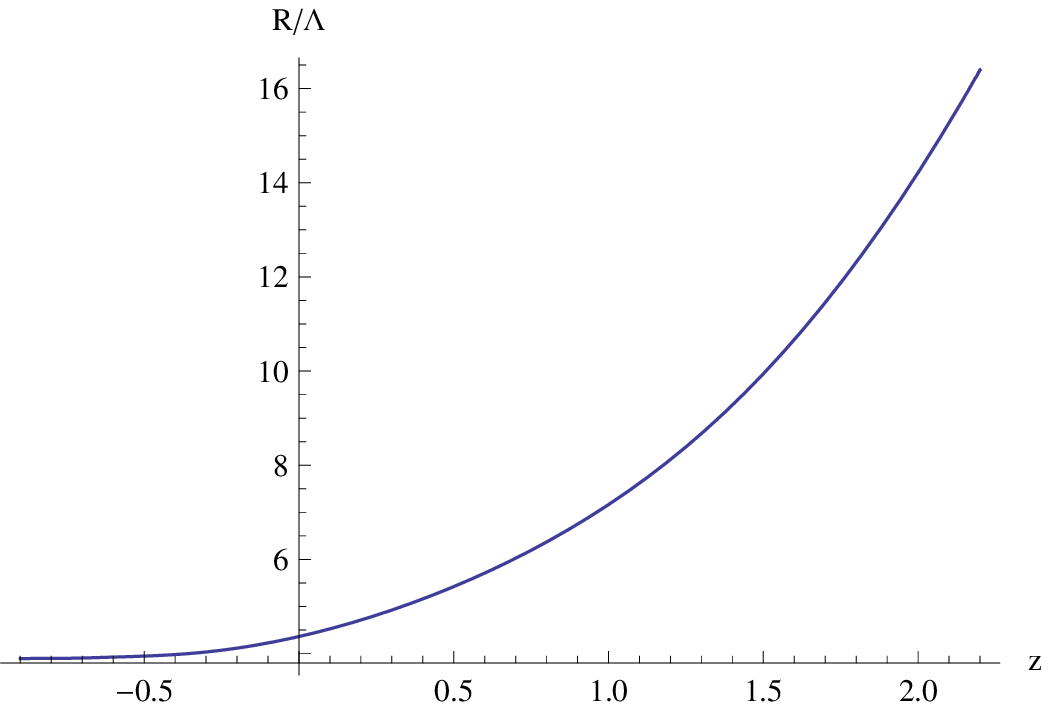}
\end{center}
\caption{Plot of $R/\Lambda$ for $R_0=0.8\Lambda$.
\label{Fig9}}
\end{figure}

\begin{figure}[-!h]
\begin{center}
\includegraphics[angle=0, width=0.5\textwidth]{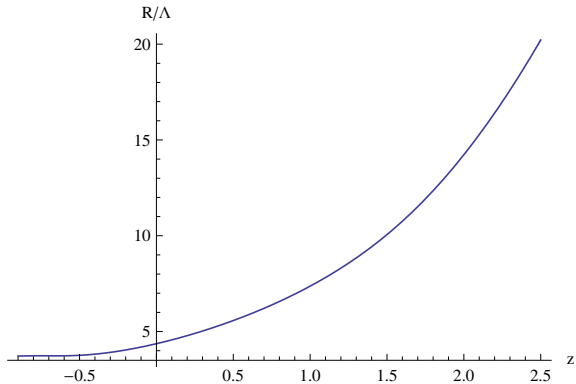}
\end{center}
\caption{Plot of $R/\Lambda$ for $R_0=\Lambda$.
\label{Fig10}}
\end{figure}

As last point, we want to analyze the behavior of the Ricci scalar
in Eq.~(\ref{Ricciscalar}) for $R_0=0.6\Lambda$, $0.8\Lambda$,
$\Lambda$. Results
are shown in Figs.~\ref{Fig8}, \ref{Fig9} and \ref{Fig10}.
We clearly see that the transition crossing
the phantom divide does not cause any serious problem to the accuracy
of the
cosmological evolution arising from our model. In particular,
$R(z\rightarrow -1^+)$ tends to $12\tilde{m}^2y_\mathrm{H}(z\rightarrow
-1^+)$, which
is an effective cosmological constant (note that $R_0$ is small and
we are
close to the value of the $\Lambda$CDM model, where
$12\tilde{m}^2y_\mathrm{H}=4\Lambda$). As a consequence, the de Sitter
solution is a
final attractor of our system and describes an eternal accelerating
expansion.


\setcounter{equation}{0}
\section{Asymptotic behavior \label{SectVIII}}

As a last issue, we will analyze the solutions of our model when $R$ is
very large in comparison with the de Sitter curvature $R_{\mathrm{dS}}$.
This means that Eq.~(\ref{total}) can be approximated by
\begin{equation}
F(R\rightarrow \infty)\simeq \gamma R^{\alpha}\,, \label{approx}
\end{equation}
which is proved by the fact that $\alpha>2$ and, by setting
$\gamma=R_{\mathrm{dS}}^{\alpha-1}$, one has $\gamma R^{\alpha} \gg
R$. In
order to check for solutions, we use Eq.~(\ref{rho}) and verify
Eq.~(\ref{FRW1}). A class of asymptotic solutions of the model of
Eq.~(\ref{approx}) at the limit $t\rightarrow 0^+$ is
\begin{equation}
H(t)=\frac{H_{0}}{t^{\beta}}\,,
\end{equation}
where $H_{0}$ is a large positive constant and $\beta$ a positive
parameter
so that $\beta=1$ or $\beta>1$. It follows from Eq.~(\ref{R}),
\begin{equation}
R\simeq 12\frac{H_0^2}{t^{2\beta}}.
\end{equation}
Eq.~(\ref{rho}) gets
\begin{equation}
\rho_{\mathrm{eff}}\simeq \frac{\delta}{t^{2\beta}}\,.
\end{equation}
Here, $\delta$ is a positive constant and Eq.~(\ref{FRW1}), in the
limit
$t\rightarrow 0^+$, is perfectly consistent.
This result shows that in the limit $R\rightarrow\infty$
the model exhibits a past singularity, which could be identified with
the Big Bang one. It is important to stress that this kind of solution is disconnected
from the de Sitter inflationary solution, where the term $R$ is of the
same order of $\gamma R^\alpha$ and is therefore not negligible as in Eq.~(\ref{approx}).
We may assume that, just after the Big Bang, a Planck epoch takes over where
physics is not described by GR and where quantum gravity effects are
dominant. When the universe exits from the Planck epoch, its
curvature is bound to be the characteristic curvature of inflation
and the de Sitter solution takes over.

\setcounter{equation}{0}
\section{On the stability of de Sitter space and a
realistic model without singularities \label{detailed}}

We here investigate in more detail the stability of the de Sitter solution
(or its absence) and construct another model which does not generate any singularity.
The de Sitter condition (\ref{dScondition}) can be rewritten as
\begin{equation}
\label{dS1}
0 = \frac{d}{dR}\left( \frac{F(R)}{R^2} \right) \, .
\end{equation}
Let $R=R_{\mathrm{dS}}$ be a solution of (\ref{dS1}). Then $F(R)$ has
the form
\begin{equation}
\label{dS2}
\frac{F(R)}{R^2} = f_0 + f_1 (R) \left( R - R_{\mathrm{dS}} \right)^n
\, .
\end{equation}
Here, $f_0$ is a constant, which should be positive if we require
$F(R)>0$.
We now assume that $n$ is an integer bigger or equal than $3$: $n\geq
3$.
Assume the function $f(R)$ does not vanish at $R=R_{\mathrm{dS}}$,
$f(R_{\mathrm{dS}})\neq 0$. Since $n\geq 3$, one finds
\begin{equation}
\label{dS5}
 - R_{\mathrm{dS}} + \frac{F'(R_{\mathrm{dS}})}{F''(R_{\mathrm{dS}})}
= 0\, ,
\end{equation}
which tells us that $m^2$ in (\ref{msquare}) vanishes.
Therefore, a more detailed investigation is necessary in order to
check stability.
Using the expression of $m_\sigma^2$,
\begin{equation}
\label{JGRG24}
m_\sigma^2 =\frac{3}{2}\left[\frac{R}{F'(R)}
 - \frac{4F(R)}{\left(F'(R)\right)^2}
+ \frac{1}{F''(R)}\right]\, ,
\end{equation}
one can investigate the sign of $m_\sigma^2$ in the region $R\sim
R_{\mathrm{dS}}$.
Note that the Eq.~(\ref{msquare}) is not used since
this expression is only valid at the point $R=R_{\mathrm{dS}}$.
Hence, we get
\begin{equation}
\label{dS6}
m_\sigma^2 \sim - \frac{3n(n-1)
R_{\mathrm{dS}}^2f_1(R_{\mathrm{dS}})}{2f_0^2}
\left( R - R _0\right)^{n-2} \, .
\end{equation}
Eq.~(\ref{dS6}) indicates that, when $n$ is an even integer, the de
Sitter
solution is stable provided $f_1(R_{\mathrm{dS}})<0$ but it is
unstable if
$f_1(R_{\mathrm{dS}})>0$.
On the other hand, when $n$ is an odd integer, the de Sitter solution
is always unstable.
Note, however, that when $f_1(R_{\mathrm{dS}})<0$ $\left(
f_1(R_{\mathrm{dS}})>0 \right)$,
we find $m_\sigma^2 >0$ $\left(m_\sigma^2 < 0\right)$ if
$R>R_{\mathrm{dS}}$, but
$m_\sigma^2 <0$ $\left(m_\sigma^2 > 0\right)$ if $R<R_{\mathrm{dS}}$.
Therefore when $f_1(R_{\mathrm{dS}})<0$, $R$ becomes small but when
$f_1(R_{\mathrm{dS}})>0$, $R$ becomes large. The stability condition
can thus be used to get realistic (unstable) de Sitter inflation for a
specific $F(R)$ gravity.

The following model, instead of (\ref{total}), is considered (compare
with \cite{Nojiri2010}),
\begin{equation}
\frac{F(R)}{R^2} = \frac{1}{\tilde R} - \frac{2\Lambda}{\tilde
R}\left(1-\mathrm{e}^{-\frac{\tilde R}{R_{0}}}\right)
+ \alpha {\tilde R}^{-\epsilon}\, , \quad
\tilde R \equiv \frac{R_i}{n} \left[ \left(\frac{R-
R_i}{R_i}\right)^n + 1
\right]\,. \label{total2}
\end{equation}
Here $n$ is assumed to be an odd integer $n\geq 3$.
We also assume that $R_i$, $\alpha$, and $\epsilon$ are positive
constants and that $\epsilon$ satisfies the
condition $0\leq \epsilon <1/n$. We choose $\alpha$ to be small
enough.
When $0< R\ll R_i$, we find $\tilde R \sim R$ and therefore the model
(\ref{model}) is reproduced.
When $R\sim R_i$, $F(R)/R^2$ behaves as in (\ref{dS2}), with
\begin{equation}
\label{R2A}
f_0 = \frac{n}{R_i} - \frac{2n\Lambda}{R_i} + \alpha
\left(\frac{R_i}{n}\right)^{-\epsilon}\, ,\quad
f_1(R_i) = - \left[ \frac{n}{R_i} - \frac{2n\Lambda}{R_i}
\left(1 - \mathrm{e}^{- \frac{R_i}{nR_0}}
+ \frac{R_i}{nR_0} \right) + \alpha \epsilon
\left(\frac{R_i}{n}\right)^{-\epsilon} \right]\, .
\end{equation}
Since $0<\Lambda\ll 1$ and it is assumed that $\alpha \ll \left(
\frac{n}{R_i} \right)^{1 - \epsilon}$, we find
\begin{equation}
\label{R2AA}
f_0 \sim \frac{n}{R_i} \, ,\quad
f_1(R_i) \sim - \frac{n}{R_i} <0 \, ,
\end{equation}
and, therefore, there exists a de Sitter solution $R=R_i$ and the
curvature always becomes smaller, slowly decreasing from the de Sitter point. Therefore,
no future singularity is generated. When $R\to \infty$, $F(R)$ behaves as
\begin{equation}
\label{R2B}
F(R) \sim \alpha R^{2-2n\epsilon}R_i^{\epsilon(n-1)} \, .
\end{equation}
Since $1<2-2n\epsilon\leq 2$, the singularity cannot emerge.
Using the same numerical techniques as in the above sections, one can numerically fit
this non-singular model with actual observable data coming from the dark energy epoch.

\section{Discussion \label{SectX}}

In summary, we have investigated in this paper some models corresponding to
the quite simple exponential theory of modified $F(R)$ gravity which are able
to explain the early- and late-time universe accelerations in a unified way.
The viability conditions of the models have been carefully investigated and it
has been demonstrated that the theory quite naturally complies with the local
tests as well as with the observational bounds. Moreover, the inflationary era
has been proven to be unstable and graceful exit from inflation has been established.
A numerical investigation of the dark energy epoch shows that the theory is
basically non-distinguishable from the latest observational predictions of the standard
$\Lambda$CDM model in this range. Special attention has been paid in the paper to the
occurrence of finite-time future singularities in the theory under consideration.
It has been shown that it is indeed protected against the appearance of such singularities.
Moreover, its evolution turns out to be asymptotically de Sitter (it has a late-time
de Sitter universe as an attractor of the system). Hence, the future of our universe, according
to such modified gravity, is eternal acceleration. We have also demonstrated that slight
modifications of the theory may lead to other non-singular exponential gravities with
similar predictions, what points towards a sort of stable class of well-behaved theories.

Very nice properties of exponential gravity are its extreme analytic simplicity,
as well as the noted singularity avoidance. In this respect, the theory considered
seems to be a very natural candidate for the study of cosmological perturbations
and structure formation, which are among the most basic issues of evolutional cosmology.
However, the theory remains in the class of higher-derivative gravities, which is not yet
well understood, even concerning its canonical formulation \cite{woodard}.
In this respect, the covariant perturbation theory developed in \cite{dunsby}
could presumably be applied for such investigation. This will be pursued elsewhere.

\section*{Acknowledgments \label{Ack}}

We are grateful to G. Cognola for his participation at the early stages of
this work. This research has been supported in part by the INFN (Trento)-CSIC
(Barcelona) exchange project 2010-2011, by MICINN (Spain) project FIS2006-02842,
by CPAN Consolider Ingenio Project and AGAUR (Catalonia) 2009SGR-994 (EE and SDO),
and by the Global COE Program of Nagoya University (G07) provided by the Ministry of
Education, Culture, Sports, Science \& Technology of Japan
and the JSPS Grant-in-Aid for Scientific Research (S) \# 22224003 (SN).

\appendix
\section{The Einstein frame \label{SectA}}
\setcounter{equation}{0}

$F(R)$ gravity may be rewritten in scalar-tensor or Einstein frame
form. In this case, one can present the Jordan
frame action
of modified gravity of Eq.~(\ref{action}) by introducing a scalar
field which couples to the curvature.
Of course, this is not exactly a physically-equivalent formulation, as
explained in Ref.~\cite{capo}.
However, Einstein frame formulation may be used for getting some of the
intermediate results in a simpler form (especially, when matter is
not accounted for).

Let us introduce the field $A$ into
Eq.~(\ref{action}):
\begin{equation}
I_{JF}=\frac{1}{2\kappa^{2}}\int\sqrt{-g(x)}\left[
F'(A) \, (R-A)+F(A)\right] d^{4}x\label{A}\,.
\end{equation}
Here ``$JF$'' means ``Jordan frame''. By making the variation of the
action
with respect to $A$, we have $A=R$. The scalar field $\sigma$ is
defined as
\begin{equation}
\sigma = -\ln [F'(A)]\label{relazioneprincipe}\,.
\end{equation}
Consider now the following conformal transformation of the metric,
\begin{equation}
\tilde g_{\mu\nu}=\mathrm{e}^{\sigma}g_{\mu\nu}\label{conforme}\,,
\end{equation}
for which Eq.~(\ref{A}) is invariant. By using Eq.~(\ref{conforme}),
we get the Einstein frame ($EF$) action of the scalar field $\sigma$:
\begin{equation}
I_{EF}=\frac{1}{2\kappa^{2}}\int\sqrt{-g(x)}
\left[
R-\frac{3}{2}g^{\mu\nu}\partial_{\mu}\sigma\partial_{\nu}\sigma-V(\sigma)\right]
d^{4}x
\label{Emmorroide}\,,
\end{equation}
where
\begin{equation}
V(\sigma)=\mathrm{e}^{\sigma}h(\mathrm{e}^{-\sigma})
 -\mathrm{e}^{2\sigma}F[h(\mathrm{e}^{-\sigma})]\label{brutto}\,.
\end{equation}
$h(\mathrm{e}^{-\sigma})$ is the solution of
Eq.~(\ref{relazioneprincipe}):
\begin{equation}
h(\mathrm{e}^{-\sigma})=A\label{h}\,.
\end{equation}
In order to pass to the scalar-tensor theory, we need the explicit
form of the potential $V(\sigma)$.
In principle, the result of Eq.~(\ref{h}) will be in the form of a
complicated transcendental function. However, in exponential gravity
the calculation simplifies a lot.

Thus, for the one-step model in Eq.~(\ref{model}) with $\Lambda=R_0$,
Eq.~(\ref{h}) leads to
\begin{equation}
h(\mathrm{e}^{-\sigma})=\ln\left(1-\mathrm{e}^{-\sigma}\right)^{-1/R_0}\,,
\end{equation}
and Eq.~(\ref{brutto}) is a simple transcendental equation which
yields the nice result
\begin{equation}
V(\sigma)=-\frac{1}{R_0}\mathrm{e}^{\sigma}(1-\mathrm{e}^\sigma)
\ln\left(1-\mathrm{e}^{-\sigma}\right)+2\Lambda
\mathrm{e}^{\sigma(2R_0-1)/R_0}\,.
\end{equation}



\end{document}